\documentclass[10pt]{article}
\usepackage[abbr]{harvard}

\usepackage[left=3cm, right=3cm, top=2cm, bottom=2cm]{geometry}

\usepackage{amsmath,amssymb}
\usepackage[abbr]{harvard}
\usepackage{xcolor}
\begin{document}
\title{Testing for multiple change-points in macroeconometrics: an empirical guide and recent developments\thanks{This manuscript is prepared for the second edition of \underline{Handbook of Research Methods and Applications on Empirical } \\ \underline{Macroeconomics}, edited by N. Hashimzade and M. Thornton, to be published by Edward Elgar.}}
\author{Otilia Boldea\thanks{Corresponding author, Department of Econometrics and OR, Tilburg University, Warandelaan 2, 5037 AB, Tilburg, The Netherlands, Email: o.boldea@tilburguniversity.edu}, Tilburg University and Alastair R. Hall, University of Manchester}
\maketitle

\begin{abstract}
    We review recent developments in detecting and estimating multiple change-points in time series models with exogenous and endogenous regressors, panel data models, and factor models. This review differs from others in multiple ways: (1) it focuses on inference about the change-points in \textit{slope parameters}, rather than in the mean of the dependent variable - the latter being common in the statistical literature; (2) it focuses on detecting - via sequential testing and other methods - \textit{multiple} change-points, and only discusses one change-point when methods for multiple change-points are not available; (3) it is meant as a \textit{practitioner's guide for empirical macroeconomists} first, and as a result, it focuses only on the methods derived under the most general assumptions relevant to macroeconomic applications.\\


\end{abstract}
\newpage
\section{Introduction}
Parametric models for macroeconomic time series involve the assumption that the specified relationship remains the same over the entire sample, a property referred to as \textit{structural stability}. However, since the earliest days of macroeconometric analysis, researchers have been concerned about the appropriateness of the assumption that such a property holds over long periods of time; for example see \citeasnoun{Tinbergen:1939}. This concern is also central to the so-called Lucas (1976) critique which has played a central role in shaping macroeconometric analysis. \citeasnoun{Lucas:1976} emphasizes the fact that the decision models of economic agents are hard to describe in terms of stable parameterizations, simply because changes in policy may change these decision models and their respective parametrization. These arguments underscore the importance of using tests of structural stability as diagnostic checks for macroeconometric models.

A large body of empirical macroeconomic studies provides evidence for parameter change in a variety of macroeconomic models. For example, there is compelling evidence that the New Keynesian Phillips curve flattened or changed over time - \citeasnoun{Hall/Han/Boldea:2012}, \citeasnoun{Magnusson/Mavroeidis:2014}, \citeasnoun{Antoine/Boldea:2018}, \citeasnoun{Inoue/Rossi/Wang:2024}, \citeasnoun{Antoine/Boldea/Zaccaria:2024}, \citeasnoun{Jorgensen/Lansing:2025}. Similarly, there is evidence that the relationship between policy interest rates and inflation/output changed over time, shifting from output targeting to inflation targeting, or to dual output and inflation targeting, and to a zero lower bound regime more recently - \citeasnoun{Schorfheide:2005}, \citeasnoun{Boivin/Giannoni:2006}, \citeasnoun{Liu/Waggoner/Zha:2009}, \citeasnoun{Hamilton:2016}, \citeasnoun{Chang/Maih/Tan:2021}. There is also strong evidence that the response of many key macroeconomic variables to
fiscal and monetary policies changes over the sample and over the state of the economy - see e.g. \citeasnoun{Auerbach/Gorodnichenko:2012}, \citeasnoun{Owyang/Ramey/Zubairy:2013}, \citeasnoun{Jorda:2020},  \citeasnoun{Cloyne:2023}, \citeasnoun{Jo/zubairy:2025}. Further examples of parameters changing over time can be found for growth, output, exchange rate and unemployment models.\footnote{See e.g. \citeasnoun{Boldea/Hall:2013a} for examples.} If these parameter changes are ignored in the estimation procedure, they lead to incorrect policy recommendations and flawed macroeconomic forecasts.

Thus, it is essential - and it has become common practice - to test parameter instability in macroeconomic models. Parameter changes can be of many types. In this chapter, we describe tests for {\it discrete} change that is, scenarios in which the parameters change abruptly at particular points in time. These points tend to be known as \textit{break-points} in the econometrics literature and as \textit{change-points} in the statistics literature; and we use these terms interchangeably.

In some cases, researchers are interested in testing whether change is associated with specific events. For example, \citeasnoun{Clarida/Gali/Gertler:2000} investigate  whether monetary policy policy reaction function of the Federal Reserve Board is different during the tenure of different chairmen. In such circumstances the break-points are specified as part of the test, and said to be \textit{known} (or fixed). While there is an obvious appeal to addressing this type of question, the outcome of such tests must be interpreted with caution because a significant statistic may reflect parameter change at a \textit{different} date. For this reason, it has become common to treat the break-points as \textit{unknown} that is, the dates of potential change are not specified as part of the test. Furthermore, there are typically many events that potentially cause parameter change and so the ideal statistical framework encompasses the case in which neither the number nor the location (in time) of the break-points is known. To date, this ideal has been developed in the context of several linear model types employed in macroeconometrics, such as univariate and multivariate regressions, panel data regressions and linear factor models. In these settings, the number and location of these change-points can be estimated via sequential testing procedures.\footnote{Note that these models are linear only in-between change-points, as change-points induce an abrupt change that can be perceived as a nonlinearity.} For nonlinear models, results are more limited, being confined in the most general scenarios - with endogenous regressors - to the case of a single break-point. 

 In linear models, a sequential testing strategy is computationally attractive because it can be performed using the dynamic programming algorithm developed by \citeasnoun{Bai/Perron:2003}. However, this inferential approach does not deliver the correct number of change-points as the sample size increases unless the significance level of each test in the procedure shrinks to zero as the sample size increases. This problem can be mitigated by controlling the size of the whole sequential procedure via a Bonferroni correction. Other inferential approaches are available that, under the right conditions, may retrieve the correct number of change-points, but may be computationally more intensive. Examples of such methods are information criteria and penalized regressions, such as adaptive lasso and mixed-integer programming. This chapter contains discussion of some of these other non-test-based procedures, but does not explore them in the same depth as the sequential testing approach due to space constraints.
 
 The literature on testing and estimating multiple change-points is vast, and spans across many areas of science and social science, from econometrics and economics to finance, statistics, environmental science, information systems, machine learning, engineering, health, to mention just a few. In the last decade, several new reviews and books have been written summarizing various change-point detection methods -  see reviews by \citeasnoun{Niu:2016}, \citeasnoun{Aminik:2017}, \citeasnoun{Casini/Perron:2019}, \citeasnoun{Namoano:2019}, \citeasnoun{Troung:2020}, and the book by \citeasnoun{Horvath/Rice:2025}. This chapter differs from all these in multiple ways: (1) it focuses on inference about the change-points in \textit{slope parameters}, rather than in the mean of the dependent variable - the latter being common in the statistical literature; (2) it focuses on detecting - via sequential testing and other methods - \textit{multiple} change-points, and only discusses one change-point when methods for multiple change-points are not available; (3) it is meant as a \textit{practitioner's guide for empirical macroeconomists} first, and as a result, it focuses only on the methods derived under the most general assumptions relevant to macroeconomic applications, including autocorrelated errors when possible. However, because it is a practical guide first, we do not formalize the results with asymptotics; interested readers are referred to the cited papers and references therein, and also the recent book by \citeasnoun{Horvath/Rice:2025}, for formal proofs of various methods.

We also discuss throughout whether the procedures available deliver consistent estimators of the change-points. The latter is possible in panel data models and in general not possible in time-series models. However, in time series models, it is typically assumed that the $m$ change-points $T_1^0, \ldots T_m^0$ are located at a fixed fraction of the sample size $T$. These fractions are denoted $\tau_j^0$ and are referred to as the break fractions. The break-points are given by $T_j^0 = [T\tau_j^0], (j=1,\ldots, m)$, where $[\cdot]$ is the smallest integer value closest to $T \tau_j^0$. In contrast to breaks points $T_j^0$, the break fractions $\tau_j^0$ can also be consistently estimated in time series models, and these estimates are often the direct result of the sequential testing procedures or other methods. 

The chapter is structured as follows. Section 2 discusses sequential tests for univariate linear time series regression models.  Attention is given to models with endogenous regressors, as the latter are pervasive in macroeconomic applications. We also discuss how to figure out which parameters change, and testing for change-points in the presence of unit roots. In all cases, we focus on allowing for dynamics and/or autocorrelated errors, and also briefly discuss detecting change-points in local projection regressions. Section 3 discusses detecting multiple change-points in multivariate time series models such as vector autoregressions and high-dimensional time series models, with many dependent variables and/or regressors. Section 4 focuses on linear panel data models, and Section 5 on linear factor models. Section 6 reviews the much scarcer literature on nonlinear models with change-points.

\section{Univariate linear models}
In this section, we summarize the literature on testing and detecting multiple discrete parameter changes in  linear parametric time series models that are relevant for macroeconomic applications. Because macroeconomic models rarely feature regressors and errors that  are independent and identically distributed, or regressors that are independent of the errors, we emphasize when possible papers that deliver the  most general assumptions and also discuss gaps in the literature related to these assumptions. 
\vspace{0.1in}

\noindent \textbf{OLS}. The most widely used testing strategy for multiple breaks in linear models estimated via ordinary least-squares (OLS) is the one proposed in the seminal paper by \citeasnoun{Bai/Perron:1998}. This strategy involves three types of tests: (i) testing no breaks versus a known number of breaks; (ii) testing no breaks against an unknown number of breaks up to a fixed upper bound, and (iii) testing $\ell$ versus $\ell+1$ breaks.

The strategy for determining the number of break-points in a sample involves, as a first step, testing zero versus a known or unknown number of breaks, via tests in (i)-(ii), described below. In practice, it is common to test for a maximum of five breaks. If the null of zero breaks is not rejected, then the model is taken to be structurally stable. If the null is rejected, then this implies the existence of at least one break, and the next step is to test for one break versus two (using the tests in (iii) below). If this test does not reject, then the estimated number of breaks is one; if the test rejects, then this evidence of at least two breaks and the next step is to test for two versus three breaks. If the test does not reject, then the estimated number of breaks is two. If the test  rejects, then the process of testing for an additional break continues until either a significant statistic is obtained or the maximum number of breaks allowed has been reached. This is a simple sequential strategy for estimating the number of breaks but it involves pre-testing bias in each step. To ensure the overall size of the procedure is at most 5\%, we suggest using a Bonferroni correction: when sequentially testing for maximum $5$ breaks, use significance level 1\% for each test.

For describing these tests, consider the following univariate linear model, estimable via OLS:
\begin{equation}
\label{dgp1}
y_t\;=\;x_t'\theta_j+u_t \qquad  (t=T_{j-1}^0+1,...,T_j^0) \qquad (j=1, \ldots, m+1)
\end{equation}
where $y_t$ is a scalar dependent variable, $x_t$ is a $p \times 1$ vector of exogenous regressors, uncorrelated with $u_t$, possibly including lags of $y_t$. The number of breaks $m$ is fixed, $T_j^0=[\tau_j^0 T]$ are the true break-points (if any), $\tau_j^0$ are the true break-fractions, for $j=1, \ldots, m$, and $\tau_0^0=0, \tau_{m+1}^0=1$ by convention. \\

\noindent \textbf{(i) Tests for a fixed number of breaks.} Under the notation above, the tests for a fixed number of breaks are for the following null and alternative hypotheses:
\begin{equation}
\label{h0LS1}
H_0:\;m=0 \qquad H_1:\; m=k, \mbox{ for a fixed } k.
\end{equation}
We can rewrite the null hypothesis in terms of restricting the parameters to be the same across sub-samples:
\begin{align} 
H_0: & \;\theta_1=\theta_2=\ldots=\theta_{k+1} \label{x} \\
H_1: & \;\theta_j \neq \theta_{j+1}  \mbox{ for all } (j=1, \ldots, k).\label{wald0:k}
\end{align}

Note that under the null hypothesis, the break-points are nuisance parameters. Denote by $\tau^c=(0, \tau_1, \tau_2,\ldots, \tau_k,1)$ the break-fractions associated with any candidate partition of the sample $T^c=(0, T_1, T_2, \ldots, T_k, T)$, where $T_j=[\tau_j T]$ for $j=1,2, \ldots, k$ such that there are enough observations in each of the $k+1$ regimes: $\tau_j \geq  \epsilon$, and $\tau_j-\tau_{j-1} \geq \epsilon$, where $\epsilon$ is usually taken to be $0.15$ to allow a maximum of five breaks in the sample. The parameters $\theta_1, \ldots, \theta_{k+1}$ are estimated piece-wise by OLS in each of the $k+1$ regimes for each candidate partition. For each of these partitions, a Wald (or LM or F) statistic for testing $H_0$ versus $H_1$ is calculated. Inference is based on the maximum statistic over all partitions, known as the sup-Wald (or sup-LM or sup-F) statistic. The sup-F test - denoted $sup\mathcal{F}_T(k)$ - and the sup-Wald test - denoted $sup\mathcal{W}_T(k)$ - are the most commonly used and the focus of this chapter. The $sup\mathcal{W}_T(k)$ was analyzed in \citeasnoun{Andrews:1993} for one change-point, and it is more general as it allows for conditional heteroskedasticity and autocorrelation.\footnote{The sup-Wald test is called a scaled sup-F test in \citeasnoun{Bai/Perron:1998} and \citeasnoun{Bai/Perron:2003}.} 

The asymptotic distributions of the $sup\mathcal{F}_T(k)$ test and of the $sup\mathcal{W}_T(k)$ are non-standard as they involve taking a maximum over a sequence of random variables. The distribution of the former was derived in \citeasnoun{Bai/Perron:1998} under two sets of dependence assumptions on the data: errors are mixingale but independent of regressors, or errors are martingale differences with respect to the $\sigma$-field that contains the regressors and lagged errors. The latter allows for lagged dependent variables. These results were generalized in \citeasnoun{Perron/Qu:2006} to allow for the product of regressors and errors to be a mixingale. Under the null hypothesis of no breaks, \citeasnoun{Bai/Perron:1998} and \citeasnoun{Perron/Qu:2006} both assume that the second moments of the data (of regressors, of regressors times errors) do not change over time asymptotically. This is a slight generalization of covariance stationarity to approximate covariance stationarity in large samples, and we call this through the rest of the text ``covariance stationarity'' as opposed to ``asymptotic mean-square error stationarity'' coined in \citeasnoun{Hansen:2000}. Assuming covariance stationarity under the null hypothesis, the distribution of the $sup\mathcal{F}_T(k)$ is pivotal in the sense that it does not depend on the data, and critical values  for this test are tabulated in \citeasnoun{Bai/Perron:1998}, \citeasnoun{Bai/Perron:2003b}, with responses surfaces for p-value calculation in \citeasnoun{Hall/Sakkas:2011}. The distribution of the $sup\mathcal{W}_T(k)$ is a scaled version of the distribution of $sup\mathcal{F}_T(k)$ the form of which follows from the derivations in \citeasnoun{Bai/Perron:1998} or \citeasnoun{Perron/Qu:2006}.

The assumption of mixingales or near-epoch dependence is general enough to cover a large set of macroeconomic applications if the data does not have a unit root or has been differenced to remove it. However, the covariance stationarity assumption is more restrictive, as most macroeconomic series, such as GDP and inflation, exhibit a smaller variance after 1984, in the wake of the "Great Moderation", and possibly larger variance yet during the financial crisis starting in 2008 and lasting several years. When the data is not (asymptotically) covariance stationary under the null hypothesis of no breaks, the asymptotic distributions of the $sup\mathcal{F}_T(k)$ and of the $sup\mathcal{W}_T(k)$ are not pivotal, and require simulations or bootstrap to obtain critical values.\footnote{Note that under the alternative hypothesis however, if the breaks in the coefficients asymptotically coincide with breaks in the second moments of the data, as would be the case in a model with lagged dependent variables, then the tests retain the same asymptotic power.} For the sup-F and sup-Wald test of zero versus only one change-point, \citeasnoun{Hansen:2000} proved that a wild bootstrap on the residuals, keeping regressors including lagged dependent variables fixed, is asymptotically valid and can be used to obtain critical values and p-values for these tests.  \citeasnoun{Boldea/Cornea/Hall:2019} generalized these results to the $sup\mathcal{F}_T(k)$ and the $sup\mathcal{W}_T(k)$ tests of zero versus $k$ breaks, allowing for the marginal distribution of the regressors and the errors to change over time, abruptly or frequently, including allowing for conditional and unconditional heteroskedasticity in the errors. They assume the errors are martingale differences multiplied by a factor that can vary over time in an arbitrary but bounded way. Unlike \citeasnoun{Hansen:2000}, they allow for the regressors to be either \textit{fixed} or \textit{recursively bootstrapped} after the residuals are bootstrapped via a \textit{wild} bootstrap. The bootstrap steps are as follows:
\begin{itemize}
    \item estimate the model under the null hypothesis of no breaks, so over the full sample;
    \item calculate the residuals and multiply them with an i.i.d. (0,1) random variables that is independent of the data, to keep the residuals mean zero and replicate their variance. This random variable can be normally distributed, or it can have the Rademacher flip-sign distribution, or the Mammen distribution, which is asymmetric;
    \item to these residuals, add back the full-sample estimates times the regressors. In the case of fixed regressor bootstrap, add the full-sample coefficient estimates times the original regressors to obtain the bootstrap data. In the case of recursive bootstrap, set the initial values of the lagged dependent variables at their original sample values, and recursively construct the lagged dependent variables using the new residuals while keeping the rest of the exogenous regressors fixed at their original sample values. In both cases, a new bootstrap sample is obtained, which is then used to recalculate the values of $sup\mathcal{F}_T(k)$ and of the $sup\mathcal{W}_T(k)$;
    \item do this $B-1$ times (typically, $B=400$ or $B=1000$). This provides a sequence of the $sup\mathcal{F}_T(k)$ tests and of the $sup\mathcal{W}_T(k)$ tests, which is then used to calculate the critical values.
\end{itemize}
Note that with covariance stationarity under the null, the $sup\mathcal{W}_T(k)$ test was also pivotal under conditional heteroskedasticity, while the $sup\mathcal{F}_T(k)$ was not. This distinction is immaterial when covariance stationarity does not hold under the null; the two distributions differ but the bootstrap replicates both, so in this case the $sup\mathcal{W}_T(k)$ test is not more general than the $sup\mathcal{F}_T(k)$ test. Additionally, the $sup\mathcal{F}_T(k)$ test is maximized asymptotically at the true break fractions as shown in \citeasnoun{Bai/Perron:1998}, because the OLS break-point estimators in \citeasnoun{Bai/Perron:1998} coincide with the maximizers of the $sup\mathcal{F}_T(k)$, and their consistency is not affected by violations of covariance stationarity. Therefore, the $sup\mathcal{F}_T(k)$ may be more powerful than the $sup\mathcal{W}_T(k)$, because the latter is not necessarily maximized at the true break-fractions when covariance stationarity is violated, as shown in \citeasnoun{Antoine/Boldea/Zaccaria:2024}.\footnote{The optimality properties of the $sup\mathcal{F}_T(k)$ and $sup\mathcal{W}_T(k)$ tests are discussed in \citeasnoun{Kim/Perron:2009} for $k=1$ under covariance stationarity assumptions, but, to the best of our knowledge,  are not known for $k>1$.} Even though the tests are not pivotal, and therefore in general the bootstrap is not expected to provide asymptotic refinements, \citeasnoun{Boldea/Cornea/Hall:2019} show by simulation that the finite sample properties of the bootstrap tests are superior to the asymptotic approximations. This was observed in earlier simulations in pivotal cases as well by \textit{e.g.} \citeasnoun{Diebold/Chen:1996} and \citeasnoun{Perron/Yamamoto:2015}. Therefore, we recommend to always use the bootstrap to obtain critical values, even in cases where asymptotic critical values are available and covariance stationarity is guaranteed. Due to the dynamic programming algorithm introduced by \citeasnoun{Bai/Perron:2003}, the computation of these tests and of those below occur in $O(T^3)$ operations regardless of the number of candidate breaks, so bootstrapping these does not pose substantial computational issues. 

Before describing the second set of tests, we would like to point out an issue with the computation of the $sup\mathcal{W}_T(k)$ test that was largely neglected in the literature. Typically, when calculating the variance in the middle of the $sup\mathcal{W}_T(k)$ test, researchers include residuals estimated imposing the alternative of $k$ breaks. This is not a good practice, as simulations in \citeasnoun{Hansen:1996} and \citeasnoun{Rothfelder/Boldea:2022} for threshold models show that the test is oversized in small samples. Intuitively, the residuals estimated under the alternative are imprecise at the edge of the sample over which the candidate partitions are considered, while the full-sample residuals are more precise. For the same reason, one should conduct the bootstrap for this test imposing the null hypothesis, therefore constructing residuals under the null of no breaks, rather than under the alternative of $k$ breaks.

\vspace{0.1in}
\noindent \textbf{(ii) Tests for an unknown number of breaks}. 
The sup-Wald test against a fixed number of breaks also rejects with probability one when $T$ is large, if the true number of breaks under the alternative is $k^* \neq k$. However, if the true alternative is $H_1: \,m=k^*$, in small samples it might not have good power properties because it is not designed for this alternative. To address this issue, \citeasnoun{Bai/Perron:1998} propose a second set of tests against the alternative of an unknown number of breaks up to a maximum:

\begin{equation}
H_0:\;m=0 \qquad H_1\;: 1 \leq m \leq M, \mbox{for a fixed } M.\label{h0LS2}
\end{equation}

These tests are known as \emph{double-maximum} or $Dmax$-type tests. The idea behind these tests is to construct for each $m \in \{1, \ldots, M\}$ a $sup\mathcal{F}_T(m)$ or $sup\mathcal{W}_T(m)$-test (thus a maximum for each $m$), and then to maximize over weighted versions of these statistics to obtain a unique test statistic for the null and alternative hypotheses in \eqref{h0LS1}. Thus, inference is based on either

\begin{align}
Dmax \mathcal{F}_T &\;=\;\max_{1\le m\le M}\frac{a_m}{p} \; Sup\mathcal{F}_T(m), \qquad \mbox{ or } \\ \qquad Dmax \mathcal{W}_T &\;=\;\max_{1\le m\le M}\frac{a_m}{p} \; Sup\mathcal{W}_T(m),\label{dmaxOLS}
\end{align}

\noindent for some fixed, strictly positive weights $a_m$. The distribution of this test generalizes in a straightforward fashion by using the asymptotic or bootstrap distributions in (i), and so the regularity conditions in (i) apply.

The weights for the $Dmax \mathcal{W}_T$ test should be set larger for a certain $m$ if one believes that $m$ is more likely to be the correct number of breaks. If there is no clear \textit{a priori} belief about the true number of breaks, one can use
equal weights $a_m=1/M$, in which case the test is known as a $UDmax$ test. However, note that the scaling $p$ is used because in its absence and with equal weights $a_m$, this
test will be equivalent to testing zero against $M$ breaks, since
the critical values increase in $m$ for a fixed $p$. Since, despite the scaling, the critical values still tend to
increase with $m$, let $c(m,p,\alpha)$ be the asymptotic critical
value of the test $Sup\mathcal{W}_T(m)/p$ at significance level $100\alpha\%$.\footnote{These critical values can be found in \citeasnoun{Bai/Perron:1998}.} Then the problem of increasing critical values is alleviated by setting $a_1=1$ and $a_m = c(1,p,\alpha) / \: c(m,p,\alpha)$, and the corresponding test is called a $WDmax$ test. Note that just like the test in (i), the bootstrap in \citeasnoun{Boldea/Cornea/Hall:2019} can be employed and its validity follows directly from the arguments in \citeasnoun{Boldea/Cornea/Hall:2019}, under more general conditions than those under which the asymptotic critical values were originally derived in \citeasnoun{Bai/Perron:1998}. 

\vspace{0.1in}
\noindent \textbf
{(iii) Tests for an additional break}. The third category of tests are called \emph{sequential-F or sequential-Wald tests}, since they are sup-F or sup-Wald tests for an additional break. The null and alternative hypotheses are, for any given $\ell$:

\begin{equation}
H_0:\;m=\ell \qquad H_1\;: m=\ell+1.\label{h0LS2}
\end{equation}

A F-type test for \eqref{h0LS2} is proposed in \citeasnoun{Bai/Perron:1998} and asymptotically equivalent Wald-type tests are also justified by the arguments in their paper and in \citeasnoun{Perron/Qu:2006} for conditionally heteroskedastic and autocorrelated errors. They can also be obtained as a special case of the sequential Wald tests in \citeasnoun{Hall/Han/Boldea:2012} and \citeasnoun{Boldea/Cornea/Hall:2019}, which allow for conditional and unconditional heteroskedasticity but not for autocorrelated errors. For the Wald test, one would ideally estimate the model only under the alternative hypothesis or $\ell +1$ breaks.\footnote{For an equivalent LR test that estimates the model in \eqref{dgp1} with $\ell$, respectively $\ell+1$ breaks, see \citeasnoun{Bai:2009}.} However, for computational ease, it has become routine among practitioners to use \possessivecite{Bai/Perron:1998} approach of pre-estimating the model with $\ell$ breaks. This implies that the estimates of the $\ell$ breaks obtained as a by-product of calculating the $Sup\mathcal{F}_T(\ell)$ are imposed as if they were the true ones, and for the alternative hypothesis in \eqref{h0LS2}, evidence is maximized for exactly one additional break, occurring in only one of the $\ell+1$ sub-samples obtained by partitioning the sample with the pre-estimated $\ell$ breaks. This implies constructing in each of $\ell+1$ regimes of the null hypothesis a $Sup\mathcal{F}_T(1)$ or $Sup\mathcal{W}_T(1)$ test, and taking the maximum of these over the $\ell+1$ regimes. The asymptotic distributions can be derived under the same regularity conditions as for (i), and are available from \citeasnoun{Bai/Perron:1998} and \citeasnoun{Perron/Qu:2006}. \citeasnoun{Boldea/Cornea/Hall:2019} prove the validity of bootstrap equivalents of these tests under conditions that  do not require covariance stationarity to hold in-between the $\ell$ breaks imposed under the null hypothesis. 
\vspace{0.1in}

\noindent \textbf{2SLS.} When some regressors are endogenous, \citeasnoun{Hall/Han/Boldea:2012} show that a similar sequential procedure for finding the number of breaks in models with endogenous regressors can be developed, based on the analogous tests constructed using the 2SLS estimates. However, unlike for OLS, with 2SLS, one needs to first check whether there are any breaks in the first-stage regression slope parameters. The first-stage can have no breaks, or breaks that are common to the second-stage, or breaks that are idiosyncratic to the first-stage; for an illustration see \citeasnoun{Boldea/Hall:2013a}. In practice, one does not necessarily know which of these scenario occurs, but the OLS sequential testing strategy above can be used to obtain the number and location of breaks in the first-stage. The first stage is then estimated at the implied break-point partition by OLS, and the predicted regressors can be used to construct 2SLS estimates and therefore the $sup\mathcal{F}_T(k)$ tests, $sup\mathcal{W}_T(k)$ tests, $Dmax$-type tests, and sequential tests. \citeasnoun{Hall/Han/Boldea:2012} show that under appropriate regularity conditions the asymptotic distributions of these tests are the same as for the OLS case. These conditions include covariance stationarity in the regimes specified under the null hypothesis of the test, near-epoch dependence of the product of regressors and instruments with the errors, and strong and valid instruments.\footnote{\citeasnoun{Perron/Yamamoto:2014} generalized these results to mixingales instead of near-epoch dependent processes. As mentioned earlier, both types of dependence are general enough for most macroeconomic applications.} Because the first-stage may also exhibit breaks under the null hypotheses of these tests, \citeasnoun{Hall/Han/Boldea:2012} proposed applying the 2SLS sequential testing strategy in each sub-sample over which the reduced form is stable. However, this is inefficient in small samples as the samples over which the reduced form is stable may be too small to allow further change-point detection. Instead, \citeasnoun{Boldea/Cornea/Hall:2019} propose that after estimation of the first-stage by imposing all the breaks that occur in that first stage, one computes the 2SLS $sup\mathcal{F}_T(k)$ tests or $sup\mathcal{W}_T(k)$ tests and sequential-F or sequential-Wald tests over the entire sample. These tests are not pivotal but \citeasnoun{Boldea/Cornea/Hall:2019} prove the bootstrap validity of the fixed regressor or recursive bootstrap with residuals resampled using a wild bootstrap. As discussed for OLS, their regularity conditions allow for the second (cross-)moments of the data and the residuals to change in an arbitrary but bounded way. They also allow for fixed changes in the coefficients of the first-stage. The optimality of these procedures is, as for OLS methods, unclear, but all the tests reject with probability one for large $T$, provided the changes in coefficients are large enough to be asymptotically detectable.

In the special case where the first-stage is stable and the regressors are covariance stationary in-between the breaks in the slope parameters of the second-stage, \citeasnoun{Perron/Yamamoto:2015} prove that the break-fractions can be consistently estimated by OLS, and that the resulting OLS estimators are more efficient than their 2SLS counterparts. However, outside of this scenario,  the OLS slope estimates for each candidate partition suffer from time-varying endogeneity bias, and therefore a search over OLS candidate partitions may not deliver consistent estimates of the true break-fractions, even when the number of breaks is known. In contrast, the bootstrap versions of the 2SLS sequential testing strategy remain valid when either the reduced form is unstable or regressors are not covariance-stationary in-between the slope parameter breaks, see \citeasnoun{Boldea/Cornea/Hall:2019}.
\vspace{0.1in}

\noindent \textbf{GMM.} In a model with no breaks and more instruments than regressors, it is known that the 2SLS coefficient estimators are less efficient than their GMM counterparts. Might GMM-based tests be used to detect the number and location of change-points? The answer - at time of writing - is that such an approach has not been justified theoretically. In fact, extant results are negative in nature. \citeasnoun{Antoine/Boldea/Zaccaria:2024} show that the implicit maximizer of the sup-Wald test for parameter change is not always consistent for the break fraction, and \citeasnoun{Hall/Han/Boldea:2012} prove a similar result for the argmax of the partial-sample GMM minimand.\footnote{Partial-sample GMM estimation involves stacking the moment conditions for the pre-break and post-break samples and allows the parameters to be different in these two sub-samples; for example see \citeasnoun{Andrews:1993}. GMM-based tests for structural change in nonlinear models are discussed in Section \ref{sect:GMM}. }
Therefore, we recommend using 2SLS and the sequential bootstrap procedure of \citeasnoun{Boldea/Cornea/Hall:2019} to determine the number and location of breaks. Once these are determined, one can employ GMM estimates for the slope parameters, as shown in \citeasnoun{Antoine/Boldea/Zaccaria:2024}, and their asymptotic distribution is as if the break locations were known.
\vspace{0.1in}

\noindent \textbf{Local projections. }OLS and 2SLS are often used to estimate impulse response functions and dynamic causal effects of a policy via local projections with or without instruments - see \textit{inter alia} \citeasnoun{Jorda:2005}, \citeasnoun{Jorda:2023}, \citeasnoun{Jorda/Taylor:2025} and \citeasnoun{Inoue/Jorda/Kuersteiner:2025}. For reasonable data generating processes, the errors of these local projection regressions are autocorrelated, see \textit{e.g.} \citeasnoun{Lusompa:2023}. Can one directly apply the sequential testing procedure above to models with autocorrelated errors? The answer is yes, but only under the assumption of piece-wise covariance stationarity in-between breaks in the slope parameters, because the assumptions in \citeasnoun{Perron/Qu:2006} and \citeasnoun{Hall/Han/Boldea:2012} allow for autocorrelated errors. However, as discussed before, the assumption of piece-wise covariance stationarity is restrictive for macroeconomic applications.

Can the bootstrap be used to circumvent this issue? \citeasnoun{Boldea/Cornea/Hall:2019} allow only for martingale difference errors, therefore cannot be used at horizons where the errors are autocorrelated. Therefore, there is a gap in the literature which relates to proving bootstrap validity of the OLS and 2SLS sequential tests for multiple change-points when the errors are autocorrelated. There are several types of bootstrap that one can employ, and we conjecture that resampling the residuals by the moving block bootstrap should ensure bootstrap validity of the OLS and 2SLS sequential testing procedures described above. However, we recommend in that case using the sequential procedure based on sup-F tests and not sup-Wald tests, because the latter - as defined in the existing literature - relies on computing HAC estimators in segments of the data, and these estimators will be very imprecise in finite samples at the boundary of the break-point partitions employed. 

Also note that if breaks are found, they may call into question whether the estimated slope parameters are indeed the dynamic causal effects researchers are after, as the proof that local projections recover the true dynamic causal effects rests on linearity and covariance stationarity over the full sample - see \citeasnoun{Stock/Watson:2018}.\\[0.1in]

\textbf{Other approaches to break estimation.} Besides testing, two other generic approaches have been proposed for estimation of the location and number of the break-points: (i) information criteria; (ii) lasso and mixed integer programming. We briefly summarize extant results for each approach in turn.

\textit{(i) Information criteria:} \citeasnoun{Yao:1988} proposed a BIC criterion for multiple changes in the mean of a variable, and proved that it consistently estimates the true number of breaks. \citeasnoun{Liu/Wu/Zidek:1997} proved that a modified BIC, with a larger penalty, correctly retrieves the number of breaks asymptotically. This result is derived under regularity conditions that restrict the regressors to be  strictly stationary ergodic processes independent of the errors, and the errors themselves to be an i.i.d. process. Under similarly strong assumptions, \citeasnoun{Hancock:2008} proved that minimizing the minimum description length criterion in \citeasnoun{Davis/Lee/Rodriguez-Yam:2006} - a criterion that employs likelihood methods to estimate simultaneously the model coefficients and the variance of the residuals -  is consistent for the true number of breaks. Following derivations in \citeasnoun{Nimomiya:2005}, \citeasnoun{Kurozumi/Tuvaandorj:2011} derived the exact penalty for multiple break models with magnitude of breaks shrinking to zero asymptotically, and showed that with exogenous regressors, minimizing a modified BIC (with a penalty on the number of breaks three times as large for estimating the breaks than for estimating a coefficient under further regularity assumptions) yields a consistent estimator of the number of breaks. \citeasnoun{Hall/Osborn/Sakkas:2015} proved a similar result for models with endogenous regressors estimated via 2SLS in a way that accounts for instability in the first-stage regression. An earlier paper by \citeasnoun{Hall/Osborn/Sakkas:2013} provided simulation results that show that with this larger penalty, the OLS-modified information criteria work better than the sequential sup-Wald test in the presence of autocorrelated errors. Given this, it seems surprising that these modified information criteria are not being employed more often by applied researchers. One reason may be that \citeasnoun{Bai/Perron:2006} showed by simulations that the criteria by \citeasnoun{Yao:1988} and \citeasnoun{Liu/Wu/Zidek:1997} are outperformed by sequential testing. However, at the time, results on the modified BICs were not fully available beyond the derivation in \citeasnoun{Nimomiya:2005}. Another reason may be that even though \citeasnoun{Hall/Osborn/Sakkas:2013} employ the modified BIC in the presence of autocorrelated errors, a proof of its consistency for autocorrelated errors is not available; however, we conjecture that the modified BIC is consistent both for OLS and 2SLS using the penalties in \citeasnoun{Kurozumi/Tuvaandorj:2011} and  \citeasnoun{Hall/Osborn/Sakkas:2017} respectively, even in the presence of autocorrelation or changes in the second moment of the data. A third reason may be that the penalties in both these papers were derived under shrinking magnitude of breaks (moderate size), and may not be strong enough when the size of the break is fixed (large). 
\vspace{0.1in}

\noindent \textit{(ii) Lasso and mixed integer programming.}  While a host of lasso methods have been proposed in statistics to estimate the number and location of change-points, the assumptions made are often not sufficiently general to cover most macroeconomic applications. \citeasnoun{Qian/Su:2016a} propose such a method for OLS and prove that the OLS-based adaptive lasso selects at least as many breaks as the true number of breaks. For the true breaks retrieved, they show that the break-fraction are consistent, under fairly general conditions on the data generating process, allowing for autocorrelated errors.
An alternative penalized least-squares procedure - based on mixed integer programming - was proposed in \citeasnoun{Prokhorov:2024}, who proved that this method delivers consistent estimates of the number of breaks and break-fractions even if there are many breaks (formally, as the number of breaks increases with the sample size). Both methods feature tuning parameters, but the latter appears to be more flexible than adaptive lasso, as it allows to enforce directly constraints such as a minimum segment size. Additionally, the mixed-integer programming is computationally more efficient than adaptive lasso. 

There are at least two cases when the adaptive lasso or the mixed integer programming procedure may be useful for macroeconomic applications. One, when breaks close to each other are of direct interest to empirical researchers, and one is not interested in estimating the parameters. A second case is the detection of change-points at the edge of the sample. In contrast, sequential testing methods are not well suited to detect either of these scenarios because of the fixed ``wedge'' ($\epsilon$ above) between regimes imposed in the construction of the data partitions.
\vspace{0.1in}

\noindent \textbf{Partial structural change.} It is possible that not all parameters change at the same time. If one would know which parameters change and which do not, or which parameters are constant over particular sub-samples, for models with exogenous regressors, the sequential testing strategy can be used by just imposing restrictions on coefficients, as shown in \citeasnoun{Bai/Perron:1998} or \citeasnoun{Perron/Qu:2006}. However, in general, one does not know which parameters change and when. If some parameters do not change over some sub-samples, this may be informative for applied work, as it would deliver more efficient parameter estimates. However, in practice, researchers rarely attempt to figure out which parameters change when. One approach to do so is the following simple two-step strategy. First, find the number and location of change-points  assuming that all coefficients change at the same time. Because the break-fractions converge faster than the parameter estimates, they can be treated as known in the next step, where an information criterion like BIC can be employed to select over all possible configurations of which parameters change and when. Because often only a small number of breaks is allowed, this is computationally straightforward and should retrieve asymptotically the true configuration of partial structural change.

\vspace{0.1in}

\noindent \textbf{Unit roots.} Unit roots and structural breaks are two types of non-stationarity that can be confounded, as shown in \citeasnoun{Perron:1989}. While there are many papers for testing for unit roots in the presence of breaks, there are much fewer papers that deal with testing for breaks in the presence of unit roots. \citeasnoun{Hansen:2000} demonstrated that the fixed regressor wild bootstrap is valid for the $sup\mathcal{F}_T(1)$ test and other related tests of zero versus one breaks even when there is a unit root. While his proof contained an error, \citeasnoun{Georgiev/Harvey/Leybourne/Taylor:2018} establish that this bootstrap method is indeed valid. Both papers assume martingale difference errors, however the latter allows for unconditional heteroskedasticity, and is derived in the context of predictive regression. The asymptotic properties of a sequential testing procedure for breaks in the presence of unit roots is to our knowledge not yet available in the literature. 

\section{Linear multivariate and high dimensional time series models}

\textbf{Multivariate models.} In principle, in multivariate linear time series models, where the dependent variable is a vector rather than a scalar, one can find the number and location of change-points equation-by-equation, by the procedures described in the previous section. After all, the usual multivariate time-series linear models employed in macroeconometrics are vector autoregressions (VARs), and those are typically estimated equation-by-equation. However, this leads to efficiency losses in the break-fraction estimates if the breaks are common across equations. 

One of the most general procedures for efficient estimation and for testing for multiple change-points in multivariate time series models of fixed dimension is the seminal paper by \citeasnoun{Qu/Perron:2007}.\footnote{See references in \citeasnoun{Qu/Perron:2007} for earlier contributions.} \citeasnoun{Qu/Perron:2007} show that under fairly general assumptions a quasi-likelihood procedure can be employed to consistently estimate the number of breaks, break-fractions, model parameters, and changes in the variance of the errors. Theses assumptions include strong-mixing regressors and errors, and other breaks in the second moments of the data. Additionally, their framework allows for breaks in different equations to differ, and these breaks across equations can also be close to each other and need not be separated by a positive fraction of the sample as in the case for univariate models. Moreover, because of employing the normal likelihood, the slope parameter estimators and the variance estimators are asymptotically independent, and therefore estimating changes in the slope parameters can be done separately from the estimation of changes in the variances of the errors. This is an important result as it allows \citeasnoun{Qu/Perron:2007} to provide an efficient dynamic programming algorithm to estimate each. 

However, before estimation, one needs to know the number of breaks in the slope parameters and in the variances of the errors. To develop a sequential-F testing procedure for estimating the number of breaks in the slope parameters, \citeasnoun{Qu/Perron:2007} assume no breaks in the variance of the errors. Hence, maximizing the likelihood ratio for a given partition and number of breaks is equivalent to maximizing a feasible generalized least-squares criterion over each candidate partition and number of breaks. Under similar conditions as for the estimation but assuming martingale difference errors and no fixed size changes in the second moments of the data, \citeasnoun{Qu/Perron:2007} derive the asymptotic distributions of the tests which are analogous to the ones for the univariate case.  A sequential testing procedure for estimating the number and location of breaks in the case the second moments of the data have breaks is available from the comprehensive treatment of various hypotheses in \citeasnoun{Perron/Yamamoto/Zhou:2020}. However, it involves further pre-testing and possibly breaking up the sample into segments with no variance changes, making it undesirable for small and moderate sample empirical applications. However, given extant results for univariate models \cite{Boldea/Cornea/Hall:2019} and for multivariate models \cite{Preuss/Puchstein/Dette:2015}\footnote{\citeasnoun{Preuss/Puchstein/Dette:2015} develop a non-parametric procedure to find the number and location of changes.}, we conjecture that a bootstrap equivalent of the \citeasnoun{Qu/Perron:2007} sequential procedure, without further pre-testing, is valid under appropriate regularity conditions. 

Once can also consistently estimate the number of change-points via the BIC information criterion in \citeasnoun{Bai:2000}. However, as discussed for univariate models, the penalty may not be strong enough to detect the correct number of breaks in finite samples, and results on the right penalty for multivariate models with change-points are not currently available to the best of our knowledge. 

\vspace{0.1in}

\noindent \textbf{High-dimensional time series models.} The literature for testing and estimating one change-point in high-dimensional time series regressions is a large and active area of research - see the recent review by \citeasnoun{Liu/Zhang/Liu:2022}. Most of it was developed in the last few years in statistics, where, for historical reasons, it is common to employ versions of a cumulative sum (CUSUM) test that examine whether sub-sample means of the dependent variables differ from their full sample means. Therefore, most tests are able to detect a change in the unconditional means of the dependent variables. While changes in slope parameters of a regression model do typically result in changes in the unconditional means of the dependent variables, this is not the case when regressors are demeaned. The asymptotic properties of CUSUM tests for zero versus one change-point are known under sparsity assumptions, and algorithms to implement these tests and to estimate the change-point are available - see  \citeasnoun{Zhao:2022} for results on such a CUSUM test for time series models, under strictly stationary and strong-mixing errors. However, the literature for (sequentially) testing for multiple change-points is less developed. \citeasnoun{Cho:2015} propose a method but for models with scalar coefficients in each equation. \citeasnoun{Liu/Zhang/Liu:2022} describe another potential procedure for multiple change-points and explain that more results are necessary to figure out the asymptotic properties of the proposed procedure. 

In high-dimensional time series models, such as VARs, the number of parameters typically grows with the dimension of the dependent variable and so sparsity assumptions are usually necessary to ensure that the test is well-behaved. Therefore, other algorithms than sequential testing, that deal with sparsity directly, have been proposed to detect the number and location of change-points, and most of them are based on variants of Lasso. Unfortunately, very few of these apply to time-series models such as high-dimensional VARs or network autoregressions, due to i.i.d. assumptions on the regressors, or assuming the regressors and errors are independent, both of which do not apply to VARs or network autoregressions. One notable exception is \citeasnoun{Bai/Safikhani:2023},\footnote{See references in \citeasnoun{Bai/Safikhani:2023} and \cite{Horvath/Rice:2025} for other exceptions.} who provide a three-step algorithm to estimate the number of change-points, their location and the parameters under sparsity assumptions. For the breaks retrieved, they show that the break-fraction estimates can be consistent if the dimension of the time-series grows slower than the size of the break magnitudes squared, so for sufficiently low dimensions and sufficiently large breaks. They also show that the resulting parameter estimates are consistent. 

\section{Linear panel data models}
\noindent \textbf{Homogeneous panels.} Panel data models relevant to macroeconomic applications typically have (at least) two dimensions: a cross-section dimension and a time dimension -  spanning individuals, firms, regions, or countries over time. In contrast to multivariate time series models, where the coefficients are different across equations, in \textit{homogeneous} panels they are assumed common across cross-sectional units. As a result, when the cross-section dimension is large, parameters can be estimated consistently at each point in time under regularity conditions. This also means that break-detection is easier, that breaks need not be separated by a fixed fraction of the sample, and that, unlike in the time-series case where only the break-fractions are consistently estimable, here the break-points are also consistently estimable. Consider the following panel data model with multiple structural changes:
$$
     y_{it}\;=\;x_{it}'\theta_j+u_{it} \qquad  (t=T_{j-1}^0+1,...,T_j^0) \qquad (j=1, \ldots, m+1)
$$
where $y_{it}$ is a scalar dependent variable, $x_{it}$ is a $p \times 1$ vector of regressors, for now excluding lagged dependent variables, $m$ is the number of breaks and $T_j^0$ are the true break-points (if any), where by convention $T_0^0=0$ and $T_{m+1}^0=T$,  $i=1,\ldots N$, and $t=1,\ldots T$.
The error has an unobserved components structure
$$
u_{it}\;=\;c_{it}\,+\,\epsilon_{it},
$$
where $c_{it}$ is potentially correlated with $x_{it}$ but $\epsilon_{it}$, the idiosyncratic error, is typically not. There are at least two popular specifications for $c_{it}$. First, $c_{it}=c_i$, referred to as a fixed-effect because it is individual specific but constant over time. Second, $c_{it}=\lambda_i^\prime f_t$, 
where $f_t$ are vectors of common factors or shocks and $\lambda_i$ is the vector of associated loadings; in this case, $c_{it}$ is referred to as an interactive fixed-effect. Fixed-effects are just special cases of interactive fixed-effects, when $\lambda_i=c_i$ and $f_t=1$.

Even in the absence of breaks, the model is not consistently estimable by OLS due to endogeneity bias arising from correlation between $x_{it}$ and $c_{it}$ when $T$ is fixed. However, it is possible to circumvent this problem by basing estimation on a transformed version of the model. For example, in the fixed-effect case, one possible transformation is the removal from the data $y_{it}, x_{it}$ of their respective means over time (commonly known as demeaning).\footnote{Other options include first-differencing the data.} For the interactive fixed-effects case, one possible transformation is to project out cross-sectional averages from the data, an approach known as the CCE (common correlated effects) and proposed by \citeasnoun{Pesaran:2006}.\footnote{For an alternative approach, see \citeasnoun{Bai:2009}.}  

\citeasnoun{Karavias/Narayan/Westerlund:2023} and \citeasnoun{Ditzen/Karavias/Westerlund:2024} show that some tests in \citeasnoun{Bai/Perron:1998} can be extended to the appropriately transformed versions of a panel data model with interactive fixed-effects. The first of these papers covers the case where $N\to\infty$ and $T$ is fixed or $T \rightarrow \infty$, with one break, and the second covers the case where both $N$ and $T$ go to infinity, with multiple breaks. \citeasnoun{Ditzen/Karavias/Westerlund:2024} provides a comprehensive treatment of a sequential testing strategy for consistent estimation of the number and location of change-points under fairly general assumptions including autocorrelated errors. They derive the asymptotic distributions of sequential sup-Wald tests, which they show are identical to the ones in \citeasnoun{Bai/Perron:1998}. They also show that the OLS estimators for multiple breaks in the CCE transformed model are consistent, and they provide an extension of the dynamic programming algorithm in \citeasnoun{Bai/Perron:2003} to panel data. The tests can be implemented via a dedicated  STATA package, see \citeasnoun{Westerlund/Karavias/Ditzen:2025b}.

The regularity conditions employed in \citeasnoun{Karavias/Narayan/Westerlund:2023} and \citeasnoun{Ditzen/Karavias/Westerlund:2024} do not, however, allow for lagged dependent variables.  It would be useful to derive the properties of a sequential procedure of estimating the number and location of breaks that allows for dynamic panels, as the results in \citeasnoun{Qian/Su:2016b}, discussed below, suggest that such a procedure may work. Additionally, \citeasnoun{Karavias/Narayan/Westerlund:2023} and \citeasnoun{Ditzen/Karavias/Westerlund:2024} do not allow the second moments of the data to change in-between slope parameter breaks, and it would be useful to develop a bootstrap equivalent of sequential procedure of estimating the number and location of breaks analogous to \citeasnoun{Ditzen/Karavias/Westerlund:2024} that would allow such changes.

Because appropriate data transformations are available for many settings, one could also consider using a simpler strategy for break-detection: employing the appropriate data transformation, obtaining consistent estimates of the slope parameters at each point in time available after the transformation, and testing via an F- or Wald- test whether adjacent parameters change. This is feasible for small $T$ with a Bonferroni correction to address multiple testing issues, but for moderate to large $T$ this would induce a large pre-testing bias.\\[0.1in] 

\noindent\textit{Other approaches:}  \citeasnoun{Qian/Su:2016b} and \citeasnoun{Li/Qian/Su:2016} propose methods based on an adaptive group-fused lasso (AGFL) penalty for multiple break-point detection in dynamic panel data models with fixed-effects and interactive fixed-effects respectively. Unlike the sequential testing procedure described above, \citeasnoun{Qian/Su:2016b} allow for dynamic panels, for fixed $T$, for many, possibly adjacent breaks, and for endogenous regressors that are correlated with both the fixed-effects $c_i$ and the idiosyncratic errors $\epsilon_{it}$. However, if the breaks can be assumed to be separated by a fixed fraction of the sample, the modified dynamic programming algorithm in \citeasnoun{Westerlund/Karavias/Ditzen:2025b} will likely be faster that the coordinate descent algorithm usually employed for adaptive lasso methods. 

To our knowledge, asymptotic results on alternative methods based on information criteria are not currently available in full generality to be applicable to dynamic panels with fixed-effects, multiple break-points and fixed $T$. \citeasnoun{Boldea/Drepper/Gan:2020} propose a BIC-type information criteria  - based on OLS  - to estimate the number and location of change-points in panel data models with fixed-effects and fixed $T$, allowing for autocorrelated errors but not for lagged dependent variables. They employ a stronger penalty for the number of breaks, and prove consistency of the estimated number and location of change-points, even when they are adjacent. Once these are obtained, a data-transformation can be employed in each stable segment to estimate the parameters of interest. However, because OLS estimators are inconsistent in the presence of fixed-effects for small $T$, like \citeasnoun{Perron/Yamamoto:2015}, they assume that the endogeneity bias is not changing, only at the same time as the slope parameters. For the cases where it changes at other parts of the sample, \citeasnoun{Boldea/Drepper/Gan:2020} propose pre-tests for these changes, to check whether the breaks retrieved are in the parameters of interest or in other parts of the model. It would be interesting to develop information criteria that directly retrieve only the breaks in the slope parameters.
\vspace*{0.1in}

\noindent \textbf{Heterogeneous panels.} In some circumstances, the assumption that the slope coefficients are homogeneous across $i$  may be considered restrictive. Various procedures are available for heterogeneous panels with one or multiple structural changes, where the coefficients are assumed to be heterogeneous across individuals or groups of individuals. To our knowledge, a sequential testing approach to infer multiple breaks has not been proposed in this setting, and, for completeness, we provide a brief summary of other available methods.

\citeasnoun{Baltagi/Feng/Kao:2016} consider a model with interactive fixed-effects, slope parameters that are heterogeneous at the individual level, and one break, estimated via CCE combined with a search for a candidate break-point. Assuming autocorrelated errors and dynamics in $e_{it}$ but precluding lagged dependent variables, they show that the break-point estimator is consistent, and they also discuss multiple breaks, but not how to estimate the number of breaks. \citeasnoun{Wang/Hu:2024} consider the case in which AGFL is applied to demeaned data and establish that this approach also yields consistent estimates of the number and location of breaks when $N,T \rightarrow \infty$, allowing for individual-level slope heterogeneity and autocorrelated errors, but not for lagged dependent variables.\footnote{See also \citeasnoun{Wang/Phillips/Su:2024} for an alternative method in the case of interactive fixed-effects.} 

Other methods for break detection have been proposed for the case where the heterogeneity of the model parameters is at the group level and group membership is unknown. \citeasnoun{Lumsdaine/Okui/Wang:2023} propose an OLS-based procedure that allows for multiple breaks separated by fixed fractions of the sample and estimates both breaks and group membership, however assuming that there are only group level fixed-effects, not individual specific fixed-effects. They show that the resulting break-point estimators are consistent and propose a BIC-type information criterion for estimating the number of breaks.  \citeasnoun{Okui/Wang:2021} propose AGFL methods for estimating multiple breaks in panel data models with group or individual level fixed-effects and group-level slope heterogeneity. They apply AGFL to first-differenced data and prove that this approach retrieves the true number of breaks and the break locations for each group with probability one in the limit as $N,T \rightarrow \infty$ under high level assumptions on dependence. They also propose information criteria to retrieve the correct number of groups and to pick the lasso tuning parameter. 

For both homogeneous and heterogeneous panels with fixed-effects, \citeasnoun{Kaddoura:2025} allows for partial structural change, and shows how to retrieve coefficient-specific breaks by penalizing the consecutive difference in each slope parameter separately, rather than jointly as in the AGFL penalty. \citeasnoun{Kaddoura:2025} proves that this adaptive fused lasso method correctly retrieves the number and location of change-points under fairly general assumptions, allowing for fixed-effects and autocorrelated errors, for $N \rightarrow \infty$ with fixed $T$ in the homogeneous case, and with large $T$ in the heterogeneous case. While \citeasnoun{Kaddoura:2025} does not allow for dynamic panels, it is likely possible to extend his method to dynamic panels, for example, using a penalized GMM criterion similar to \citeasnoun{Qian/Su:2016b}.

\section{Factor models and factor-augmented forecasting equations}
\label{sect:factor}
With the advent of big-data, it has become popular to use large-scale approximate factor models as a method of dimension reduction in the analysis of macroeconomic systems. One popular use of the estimated factors is in so-called factor-augmented forecasting equations. This section reviews methods for testing for structural change in both contexts.\\[0.1in]

\noindent\textbf{Factor models.} Large-scale factor approximate models involve the assumption that a $N\times 1$ vector of variables $x_t$ are observed at $T$ time periods and $x_t$ is generated via the following process
\begin{equation}
  \label{eq:fmss}  
x_t\;=\;\Lambda f_t\,+\,e_t,\qquad t=1,2,\ldots,T,
\end{equation}
where $f_t$ is a $r\times 1$ vector of latent factors, $\Lambda$ is a $N\times r$ matrix of unknown factor loadings and $e_t$ is a $N\times 1$ vector of idiosyncratic errors. Within this framework, the systematic component of $x_t$, $\Lambda f_t$, is driven by  the $r$ factors where it is assumed that $r<<N$. Here ``large-scale'' refers to the fact that both $N$ and $T$ go to infinity in the statistical analysis of the model. The adjective ``approximate'' refers to a collection of regularity conditions on $f_t$ and $e_t$ that allow \textit{inter alia} for $f_t$ to be a serially correlated process, for the idiosyncratic errors to exhibit mild forms of cross-sectional and serial dependence, and for the factors and errors to be weakly dependent. It is customary to estimate this model via the method of principal components (PC). In most applications $r$ is unknown and it has become popular to estimate $r$ using a member of the class of the information criteria-based methods that have been shown to be consistent by \citeasnoun{Bai/Ng:2002}; hereafter referred to as BN's method. \citeasnoun{Bai:2003} provides the large sample distribution theory for the PC estimators of both loadings and factors in the model where $r$ is known. Due to a fundamental lack of identification inherent in the model, the estimated factors and loadings are only consistent up to a rotation, and when we use the term consistency in this context below the qualifier ``up to a rotation'' is to be taken as implicit. We refer the reader to Chapter 11 of this handbook for a review of factor models and a description of PC.

Within this framework,  the term ``structural instability/change'' is commonly used to describe the case in which the loadings matrix changes over time and we follow that practice here. All the tests described below involve testing against discrete change in the loadings at one or multiple break-points using statistics based on PC estimators of the factors. To facilitate the discussion the tests, it is useful to outline the properties of the PC estimators in factor models with neglected structural instability of this kind.

To this end, consider the case in which
\begin{equation}
  \label{eq:fmsu}  
x_t\;=\;\Lambda_t(\tau) f_t\,+\,e_t,\qquad t=1,2,\ldots,T,
\end{equation}
where
\begin{equation}
   \label{eq:lambdass}
   \Lambda_t(\tau)\;=\; \Lambda_1 \mathcal{I}(t\leq [\tau T])\,+\,\Lambda_2\mathcal{I}(t>[\tau T]), 
\end{equation}
and $\mathcal{I}(\cdot)$ is an indicator function. Since $\Lambda_t(\tau)=\Lambda_1 +\xi_t$ for $\xi_t=\mathcal{I}(t>[\tau T])(\Lambda_2-\Lambda_1)$, the model in \eqref{eq:fmsu} can be written as
$$
x_t\;=\;\Lambda_1 f_t\,+\,a_t,\qquad t=1,2,\ldots,T,
$$
where $a_t=e_t+\xi_tf_t$. Thus $x_t$ has a factor model representation with constant loadings and the neglected structural instability as part of the composite error term. \citeasnoun{Batesetal:2013} show that if the neglected instability is ``small'' then $f_t$ and $a_t$ satisfy the conditions for the approximate factor model under which the PC estimators are consistent.\footnote{\citeasnoun{Batesetal:2013} provide generic conditions under which the PC-estimated factors deliver a consistent estimator of the space spanned by the true factors and BN's method provides a consistent estimator of $r$. Also see \citeasnoun{Stock/Watson:2002a}.}  Given  instability of the form in \eqref{eq:lambdass}, ``small'' can be interpreted in terms of the fraction of the series whose loadings actually change by a non-zero $O(1)$ amount at the break-point. \citeasnoun{Batesetal:2013} show that if this fraction is $O(N^{-1/2})$ then the neglected structural change is ``small'' in the sense defined above, and that this fraction needs to be $O(1/min(N,T))$ for the consistency of BN's method, a stronger condition in most macroeconomic applications.\footnote{For example, at time of writing, the FRED Quarterly and Monthly data sets have respectively $(N,T)$ approximately equal to $(245, 270)$ and $(130,800)$, implying very small changes in loadings. For further information on these data sets see \citeasnoun{McCracken/Ng:2016}.}

If the neglected structural instability is not ``small'' then \citeasnoun{Breitung/Eickmeier:2011} observe that $x_t$ has a factor model representation with constant loadings involving $q$ factors where $r\leq q\leq 2r$ that is,
\begin{equation}
    \label{eq:fmsu_mat}
    X\;=\;G\Theta^\prime\,+\,e
\end{equation}
where $X$ is the $T\times N$ matrix with $t^{th}$ row $x_t^\prime$, $e$ is the $T\times N$ matrix with $t^{th}$ row $e_t^\prime$, and $G$ and $\Theta$ are respectively the $T\times q$ and $q \times N$ matrices. To emphasize the difference between $G$ and $F$, it has become common to refer to $G$ as the \textit{pseudo} factors and $F$ as the \textit{true} factors. In general, only in the case where there is no structural change that is, $\Lambda_1=\Lambda_2=\Lambda$, do the pseudo and true factors coincide because 
then $x_t$ is generated by \eqref{eq:fmsu_mat} with $q=r$, $G=F$, and $\Theta=\Lambda$.

\citeasnoun{Han/Inoue:2015} show if the model is structurally unstable then PC consistently estimates the components $G$ and $\Theta$ in  \eqref{eq:fmsu_mat}, and further that  BN's method is consistent for $q$ in certain leading cases of empirical relevance, such as when there are emerging or disappearing factors or when loading matrices before and after the break are linearly unrelated.\footnote{Similar results are presented in \citeasnoun{Chenetal:2014} under more restrictive conditions.}

With this background, we turn to summarizing methods proposed for testing structural stability in factor models against an alternative with abrupt breaks. In our discussion, we group these methods into three categories: (i) tests for breaks in an individual series at a single break-point; (ii) tests for common structural change in multiple series at a single break-point; (iii) tests for coincident structural change in multiple series at multiple break-points. While these tests differ in certain aspects, all the tests have certain features in common that, for reasons of efficiency in presentation, we highlight first.  As noted above, all tests are based on PC estimators with BN's method employed to estimate the number of factors. In addition, the asymptotic theory is premised on the condition that $\sqrt T/N\to 0$ as $T,N\to\infty$: under this condition the large sample inferences about the loadings are unaffected by the estimation error associated with the factors, see \citeasnoun{Bai:2003}. For categories (i) and (ii), known and unknown break-point versions of the tests are proposed. We focus primarily on the former as it serves to highlight the ideas behind the various tests and the differences between them. In each case, the unknown break-point versions can be constructed using the supremum of the sequence of their fixed-break-point counterparts.\\[0.1in] 

\noindent \textit{Single equation, single break:} \citeasnoun{Breitung/Eickmeier:2011} propose Wald, LR and LM statistics for testing whether there is a break in the loadings for a single variable, $x_{i,t}$ say. The null and alternative hypotheses are
$$
H_0(\tau): \Lambda_{1}\;=\;\Lambda_{2}
 \mbox { vs } H_1^{(i)}(\tau): \lambda_{1,i}\;\neq\;\lambda_{2,i},
$$
where $\lambda_{j,i}^\prime$ denotes the the $i^{th}$ row of $\Lambda_j, (j=1,2)$ defined in \eqref{eq:lambdass}.  Notice that the null involves the stability of all loadings even though the alternative is stated in terms of the loadings for $x_{i,t}$. To illustrate this type of test,  consider the regression 
$$
x_{i,t}\;=\;\hat g_t^\prime \lambda_{1,i}\mathcal{I}(t\leq [\tau T])\,+\,\hat g_t^\prime \lambda_{2,i}\mathcal{I}(t> [\tau T])\,+\,\mbox{ error}
$$
where $\hat g_t$ denote the estimated pseudo factor vector $g_t$. The LR-type test is a comparison of the unrestricted residual sum of squares from this regression with the restricted residual sum of squares obtained by imposing $\lambda_{1,i}=\lambda_{i,2}$.  Under the null hypothesis,  \citeasnoun{Breitung/Eickmeier:2011} show the test has a limiting  $\chi_r^2$ limiting distribution. However, under the alternative $q\geq r$ and $g_t\neq f_t$, and as argued above, PC estimates the model in \eqref{eq:fmsu_mat} which has \textit{constant} loadings. As a result, \citeasnoun{Yamamoto/Tanaka:2015} show that  \possessivecite{Breitung/Eickmeier:2011} tests do not reject with probability one in large samples and possess non-monotonic power functions. \citeasnoun{Yamamoto/Tanaka:2015} propose  alternative tests based on including a subset of the elements of $g_t$ in the regression model. This modified test has a non-standard limiting distribution under the null for which \citeasnoun{Yamamoto/Tanaka:2015} provide critical values.

Given that the null hypothesis involves the stability of all loadings, one concern about the single equation approach is that inferences about the stability of the loadings for $x_{i,t}$ can be contaminated by instability in the loadings for other variables in $x_t$.\\[0.1in]

\noindent\textit{Multiple equations, single break:} \citeasnoun{Han/Inoue:2015} propose tests for the joint null that all factor loadings are stable against the alternative that a fixed (non-zero) fraction of the series exhibit structural changes in the factor loadings at a common break date that is,
$$
H_0(\tau):\,\Lambda_1\;=\;\Lambda_2 \qquad \mbox{ and } H_1(\tau):\, H_1^{(i)}(\tau) \mbox{ holds  for } i\,\in\,\mathcal N,
$$
where $\mathcal N\,\subseteq\, \{1,2,\ldots,N\}$ with $|\mathcal N|=[\alpha N]$ for $\alpha\,\in\,(0,1]$.\footnote{\citeasnoun{Han/Inoue:2015} consider a more general framework in which the instability is restricted to the loadings of a subset of factors.} \citeasnoun{Chenetal:2014} refer to  $H_1(\tau)$ as involving a ``big break'' to distinguish this state of the world from the ``small break'' scenario discussed above.

\citeasnoun{Han/Inoue:2015} propose inference methods based on the Wald and LM statistics tests for testing whether the sub-sample means of $\hat g_t\hat g_t^\prime$ are equal or not at the predetermined break $[\tau T]$. These tests are based on 
$$
vech\left\{\,\{[\tau T]\}^{-1}\sum_{t=1}^{[\tau T]} \hat g_t\hat g_t^\prime\,-\,\{T-[\tau T]\}^{-1}\sum_{t=[\tau T]+1}^T \hat g_t\hat g_t^\prime\,\right\},
$$
where $vech\{\cdot\}$ is the operator that stacks the lower triangular elements of the matrix in the parentheses into a vector. \citeasnoun{Han/Inoue:2015} show that under $H_0(\tau)$ these statistics have a limiting $\chi_{r(r+1)/2}$ distribution and show their test is consistent against $H_1(\tau)$.

\citeasnoun{Chenetal:2014} consider a framework in which there may be both small and big breaks in the loadings at the break-point. Accordingly they specify the null and alternative hypotheses of interest as 
$$
H_0:\,q\;=\;r \qquad \mbox{ and } \qquad H_1:\, q\;=\;\,r\,+\,k_1
$$
where $q$ is the number of factors in \eqref{eq:fmsu_mat} and $k_1$ is some positive integer. Here $k_1$ represents the degree to which the estimated number of factors has been inflated by a ``big break'' in the loadings.\footnote{Recall that small breaks do not inflate the number of estimated factors by definition.} \possessivecite{Chenetal:2014} test is based on the regression of the first estimated factor on the remaining estimated factors that is, 
$$
\hat g_{t,1}\;=\;c^\prime \hat g_{t}^{(-1)}\,+\mbox{error} 
$$
where $\hat g_{t,1}$ is the estimator of the first element of $g_t$ and $\hat g_{t}^{(-1)}$ is the estimator of the $(q-1)\times 1$ vector containing all elements of $g_t$ apart from the first. They propose inference is based on either the Wald or LM statistics for testing test whether there is a break in $c$. Under $H_0(\tau)$, the test statistic has a limiting $\chi_{r-1}$ distribution. 

Comparing the procedures, it can be seen that both approaches are based on the information in the second moments of the estimated factors but  \citeasnoun{Chenetal:2014} test implicitly exploits only a subset of this information whereas \possessivecite{Han/Inoue:2015} test uses all the information. Simulation evidence  suggests neither test uniformly dominates in terms of power properties in settings with $q>r$. However, \citeasnoun{Han/Inoue:2015} show that their test dominates for cases with $q=r$ because \possessivecite{Chenetal:2014} test has low power against alternatives for which the number of factors is not inflated by the instability.

\citeasnoun{Baietal:2024} consider testing against a single big break at an unknown break-point. They propose a sup-LR test statistic derived under the assumption that the likelihood is Gaussian. If the likelihood function is correctly specified and the null hypothesis of structural stability holds then their test statistic is asymptotically equivalent to \possessivecite{Han/Inoue:2015} sup-Wald statistic. However, if the likelihood is misspecified then the limiting distribution of the sup-LR statistic under the null is non-standard and has to be simulated on a case-by-case basis.  While both the sup-LR and sup-Wald tests both reject with probability one in large samples, \citeasnoun{Baietal:2024} show that the sup-LR test is more powerful in the sense that it diverges faster under the alternative hypothesis than the sup-Wald. This power gain is attributable to the superior properties of the implicit break-point estimator based on the sup-LR statistic relative to its counterpart based on the sup-Wald statistic.\\[0.1in]

\noindent\textit{Multiple equations, multiple breaks:} \citeasnoun{Baltagietal:2021} propose methods for estimation and inference in factor models with multiple breaks. Letting $m$ denote the number of breakpoints, 
\citeasnoun{Baltagietal:2021} propose tests for: $H_0: m=0$ versus $H_1:m=k$; $H_0: m=0$ versus $H_1:m>0$; $H_0:m=\ell$ vs $H_1: m=\ell+1$ from a sequential testing procedure can be derived  to retrieve the number and location of the breaks. Just as with the multiple equations, single break tests above, these tests are based on the second moments of the PC-estimated factors. For example, their sup-F statistic for $H_0: m=0$ versus $H_1:m=k$ is based on a comparison of measures of the variation in $\hat g_t\hat g_t^\prime$ about its mean under the null and the alternative. Subject to certain regularity conditions, \citeasnoun{Baltagietal:2021} show that under the null hypothesis their sup-F statistic converges in distribution to the distribution of \possessivecite{Bai/Perron:1998} sup-F test for the analogous hypotheses in the linear regression model with degrees of freedom equal to $\hat q(\hat q+1)/2$. For the tests of $H_0: m=0$ versus $H_1:m>0$ and $H_0:m=\ell$ versus $H_1: m=\ell+1$, \citeasnoun{Baltagietal:2021} adapt the approaches in \citeasnoun{Bai/Perron:1998} for the analogous hypotheses to this setting to provide statistics that are functions of the second moments of the estimated factors over appropriately defined sub-samples. Once again the limiting distributions under the null hypotheses are the corresponding distributions derived by \citeasnoun{Bai/Perron:1998}, with the caveat that this only holds for the test of $H_0:m=\ell$ versus $H_1: m=\ell+1$ if the estimated number of factors is the same in all sub-samples over which the supremum is taken.\footnote{If this condition does not hold then the limiting distribution can be simulated on a case-by-case basis, see \citeasnoun{Baltagietal:2021} for further details. }

Assuming all breaks are ``big'', \citeasnoun{Baltagietal:2021} demonstrate that a sequential testing strategy yields a consistent estimator of the number of breaks provided the significance level of the tests tends to zero as $N,T\to\infty$. As they argue, the intuition behind this result is that the second moment matrix of the pseudo-factors have changes at exactly the same dates as the loadings.

The estimation of both the break-point/fractions and the number of factors in large scale approximate factor models with multiple breaks has also been addressed outside the context of testing. The interested reader is referred to \textit{inter alia} \citeasnoun{Chengetal:2016}, \citeasnoun{Baietal:2020} and \citeasnoun{Baltagietal:2021}.\\[0.1in]
 
\noindent\textbf{Factor-augmented forecasting equations.}
Following the seminal work of \citeasnoun{Stock/Watson:2002b}, an important use of estimated factors is in forecasting models of the type
\begin{equation}
\label{eq:far}
y_{t+h}\;=\;\alpha_{h,t}\,+\,\beta_{h,t}^\prime \hat g_t\,+\gamma_{h,t}^\prime x_t\,+\,\varepsilon_{t+h}
\end{equation}
where $y_{t+h}$ is the variable to be forecast, $h$ is the forecast horizon, $x_t$ is a vector of predetermined variables and $\varepsilon_{t+h}$ represents the error. \citeasnoun{Stock/Watson:2002b} refer to the estimated factors as diffusion indexes and as a result, the model in \eqref{eq:far} is often referred to as a \textit{diffusion-index forecasting equation}. In this context structural instability can be due to a change in the data generation process for the factors and/or changes in the parameters of the forecasting equation. \citeasnoun{Corradi/Swanson:2014} propose tests of parameter constancy against an alternative with potentially multiple breaks in either the loadings of the underlying factor model and/or the parameters of the forecasting equation. For ease of presentation, we focus here on the single-break case and no other variables are included in the forecasting equation apart form the intercept which is assumed to be time invariant that is, $\alpha_{h,t}=\alpha_h$ and $\gamma_{h,t}=0$.\footnote{See \citeasnoun{Corradi/Swanson:2014}[p.104] for a discussion of how to adapt their test to  allow for changes in the intercept.}  
The part of the hypotheses involving the loadings can be expressed using the components of \eqref{eq:fmsu} and \eqref{eq:lambdass}. Similarly define
$$
\beta_{h,t}\;=\;\beta_1\mathcal I\{t\leq [\nu T]\}\,+\,\beta_2 \mathcal I\{t> [\nu T]\},
$$
where $\nu\in(0,1)$. Note that the break fractions $\nu$ and $\tau$ need not coincide although they may do.
The null hypothesis is both loadings and forecast equation parameters remain constant over the sample,
$$
H_0: \Lambda_1\,=\,\Lambda_2 \mbox{ and } \beta_1\,=\,\beta_2.
$$
Under the alternative at least one of these sets of parameters changes at some point in the sample,
$$
H_1:\,\Lambda_1\,\neq\,\Lambda_2 \mbox{ and/or } \beta_1\,\neq\,\beta_2.
$$
\possessivecite{Corradi/Swanson:2014} test is based on the comparison of two estimators of the covariance between the variable to be predicted, $y_{t+h}$, and the estimated factors, $\hat g_t$: one based on the full sample and one based on a rolling window. Setting $T=P+R$, where $R$ represents the number of observations in the rolling window, and assuming $P/R\to\pi>0$ as $T\to\infty$, they show that under the null hypothesis their test statistic converges to a limiting normal distribution and establish the validity of a block bootstrap for obtaining critical values for the test.

\section{Testing for change-points in nonlinear models}
\label{sect:GMM}
\noindent
\textbf{Nonlinear regression models.} For nonlinear regression models estimated via nonlinear least squares, \citeasnoun{Boldea/Hall:2013b} develop an analogous sequential testing procedure to the one developed by \citeasnoun{Bai/Perron:1998} for linear models estimated via OLS. Assuming piece-wise geometric ergodicity of the data in-between breaks in slope parameters, and smoothness of the nonlinear regression function in-between breaks, they show that the asymptotic distributions of the sequential sup-F tests, now based on nonlinear least squares rather than OLS, are the same as in \citeasnoun{Bai/Perron:1998}, and they automatically deliver consistent estimates of the break-fractions. For nonlinear multivariate and panel data models, equivalent procedures are not currently available to the best of our knowledge.\\[0.1in]

\noindent
\textbf{Other nonlinear models.} For more general forms of nonlinearity, tests for a single break have been developed for models estimated via Generalized Method of Moments (GMM); see Chapter 14 of this volume. As befits the GMM framework, the null and alternative hypotheses are expressed in terms of the population moment condition. The null hypothesis is that the $(q\times 1)$ population moment condition holds at the same $(p \times 1)$ parameter value, $\theta_0$, throughout the sample that is,
\begin{equation}
\label{h0}
H_0:\;E[f(v_t,\theta_0)]\;=\;0,\qquad \mbox{ for }t=1,2,\ldots, T,
\end{equation}
where $v_t$ contains the random variables in the model. Two types of alternative are considered: (i) \textit{parameter change}, in which the population moment condition holds at different parameter values before and after the break; (ii) \textit{other forms of structural change}, in which the population moment condition holds at some parameter value before the break but at no parameter value after the break (or vice versa). As discussed below, the latter is only possible in the case where the parameter vector is overidentified that is, $q>p$. In both cases, known and unknown break-point versions of the tests are available. For ease of comparison, we start by considering the fixed break-point versions.\\[0.1in] 

\noindent
\textit{Parameter change:} Here the alternative hypothesis is 
$$
H_1(\tau):\;\begin{array}{l@{,\qquad}l}E[f(v_t,\theta_1)]\;=\;0&\mbox{for } t=1,2,\ldots,[\tau T],\\
E[f(v_t,\theta_2)]\;=\;0&\mbox{for } t=[\tau T]+1,\ldots,T, \end{array}
$$
where $\theta_1\neq \theta_2$ and $\tau$ is known. \citeasnoun{Andrews/Fair:1988} propose Wald, Lagrange Multiplier (LM) and Difference (D) statistics  to test $H_0$ versus $H_1(\tau)$. Subject to certain regularity conditions, all three test statistics converge to $\chi_p^2$ distribution under the null hypothesis. These regularity conditions include the crucial assumptions that $v_t$ is non-trending, the function $f(v_t,\cdot)$ is smooth in $\theta$, and that the parameters are both globally and first-order locally identified. \footnote{In instrumental variables estimation, this identification scenario equates to the case of ``strong instruments''. For break-point tests in the presence of weak instruments, see \citeasnoun{Caner:2011}.}\\[0.1in]

\noindent
\textit{Other forms of structural instability:} Here the alternative is that
$$
H_1^\prime(\tau): H_{1,a}(\tau) \mbox{ or } H_{1,b}(\tau),
$$
where
$$
H_{1,a}^\prime(\tau):\;\bigg\{\begin{array}{l@{,\qquad}l}E[f(v_t,\theta_0)]\;=\;0&\mbox{for } t=1,2,\ldots,[\tau T],\\
E[f(v_t,\theta_0)]\;\neq\;0&\mbox{for } t=[\tau T]+1,\ldots,T, \end{array}
$$
and
$$
H_{1,b}^\prime(\tau):\;\bigg\{\begin{array}{l@{,\qquad}l}E[f(v_t,\theta_0)]\;\neq\;0&\mbox{for } t=1,2,\ldots,[\tau T],\\
E[f(v_t,\theta_0)]\;=\;0&\mbox{for } t=[\tau T]+1,\ldots,T. \end{array}
$$
If $q=p$ - and so there are the same number of moment conditions as parameters - then it can be shown that $H_1(\tau)$ and $H^\prime_1(\tau)$ are equivalent.\footnote{This result presumes certain  (relatively weak) regularity conditions hold and can be established via a similar argument to \citeasnoun{Hall/Inoue:2003}[Proposition 1].} However, if $q>p$ - and so there are more moment conditions than parameters - then this equivalence does not hold. Exploiting the decomposition of the population moment condition inherent in GMM estimation,\footnote{For example, see Chapter 14 of this handbook.} \citeasnoun{Hall/Sen:1999} show that $H_1^\prime(\tau)$ can be decomposed into two parts: structural instability in the identifying restrictions at $[T\tau]$ and structural instability in the overidentifying restrictions at $[T\tau]$. Instability of the identifying restrictions is equivalent to $H_1(\tau)$ that is, to parameter change at $T\tau$. Instability of the overidentifying restrictions means that some aspect of the model beyond the parameters alone has changed at $[T\tau]$, see \citeasnoun{Hall/Sen:1999} or \citeasnoun{Hall:2005} for further discussion. 

To test against instability of the identifying restrictions,  the Wald, LM or Difference statistics for testing against $H_1(\tau)$ can be used. To test against instability of the overidentifying restrictions, \citeasnoun{Hall/Sen:1999} propose basing inference on the sum of the overidentifying restrictions test statistics from the sub-samples before and after the break.\footnote{See Chapter 14 of this handbook for a description of the overidentifying restrictions test.}  \citeasnoun{Hall/Sen:1999} show that under $H_0$ this statistic both converges to a $\chi_{2(q-p)}$ and is asymptotically independent of the statistics for testing parameter change, such as the Wald statistic. A local power analysis underscores the notion that the two tests are sensitive to different aspects of instability, and provides some guidance on the interpretation of significant statistics, although the local nature of the results needs to be kept in mind.\\[0.1in]

\noindent
\textit{Unknown break-point tests.} Above we have discussed unknown break-point tests in the context of  specific models. In each case, inference is based on the supremum of the sequence of fixed break-points indexed by $\tau$. This choice has the intuitive appeal of basing inference on the member of the sequence of fixed break-point tests that maximizes the evidence of instability.  Notwithstanding this appeal, two other functionals have also been proposed that lead to the so-called Average (``Av-'') and  Exponential (``Exp-'') statistics.\footnote{These functions can be justified as providing the test that maximizes power against a local alternative in which a weighting distribution is used to indicate the relative importance of departures from parameter constancy in different directions ( {\it i.e.} $\theta_1-\theta_2$) at different break-points and also the relative importance of different break-points; see \citeasnoun{Andrews/Ploberger:1994} and \citeasnoun{Sowell:1996a}.}  Various statistical arguments can be made to justify one statistic over another, but in practice practitioners tend to report either just the sup-test (for the reasons given above) or all three. 

Under $H_0$, the statistics above can be shown to converge weakly to functions of a $p-$ dimensional Brownian Bridge.\footnote{For regularity conditions and proofs see: Sup-test, \citeasnoun{Andrews:1993}; Av-, Exp- tests,  \citeasnoun{Andrews/Ploberger:1994} (in context of maximum likelihood) and \citeasnoun{Sowell:1996a} (in context of GMM).}  Percentiles are reported in \citeasnoun{Andrews:2003} (Sup-) and \citeasnoun{Andrews/Ploberger:1994} (Av- and Exp-).  \citeasnoun{HansenB:1997} reports response surfaces which can be used to calculate approximate p--values for all three versions of these tests.

\citeasnoun{Hall/Sen:1999} also propose unknown break-point versions of their tests based on the Sup-, Av- and Exp- functionals. Under the null hypothesis, these statistics converge weakly to the functions of a $(q-p)$ - dimensional Brownian motion; percentiles of these distributions are reported in \citeasnoun{Hall/Sen:1999}. \citeasnoun{Sen/Hall:1999} report response surfaces which can be used to calculate approximate p--values for all three versions of these tests.

To our knowledge, the extension of these methods to test for multiple breaks remains an open research question. Even if feasible, two considerations may limit their application. Firstly, nonlinear models are inherently more difficult to estimate than their linear counterparts and so the smaller sample sizes inherent in data partitioning may lead to problems with convergence for estimation routines. Secondly, even if there are no problems with convergence, it is questionable whether the sub-sample sizes with macroeconomic data would be sufficiently large for this asymptotic theory to provide a reliable guide to finite sample behaviour of the underlying statistics in nonlinear models. These problems can be circumvented by adopting  an alternative specification for the structural change.   \citeasnoun{Lietal:2024} propose a GMM estimator for the case in which the population moment condition is indexed by a parameter vector that changes smoothly through the sample. They propose tests for structural stability against an alternative of smooth parameter change and demonstrate this test also power against alternatives characterized by discrete parameter change.

\ifx\undefined\BySame
\newcommand{\BySame}{\leavevmode\rule[.5ex]{3em}{.5pt}\ }
\fi
\ifx\undefined\textsc
\newcommand{\textsc}[1]{{\sc #1}}
\newcommand{\emph}[1]{{\em #1\/}}
\let\tmpsmall\small
\renewcommand{\small}{\tmpsmall\sc}
\fi

\end{document}